\documentclass[epj]{svjour}
\usepackage{graphics}
\begin{document}
\title{Study of  semileptonic and nonleptonic decays of the $B_c^-$
meson}
\author{E. Hern\'andez\inst{1}, J. Nieves\inst{2} \and 
J.M. Verde-Velasco\inst{1}
}                     
%
%
\institute{Grupo de F\'\i sica Nuclear, Departamento de F\'\i sica Fundamental e
IUFFyM, Universidad de Salamanca, Spain. \and Departamento de F\'\i sica At\'omica,
Molecular y Nuclear, Universidad de Granada, Spain.}
\date{Received: date / Revised version: date}
%
\abstract{
We evaluate semileptonic and two--meson nonleptonic decays of the $B_c^-$ meson in the
framework of a nonrelativistic quark model. The former are done in spectator
approximation using one--body current operators at the quark level. Our model 
reproduces the constraints of  heavy quark spin symmetry  obtained in the 
limit of infinite heavy quark mass. For the two--meson nonleptonic decays
 we work  in factorization approximation.  We compare our results to the ones obtained in
different relativistic approaches.
 }
\maketitle
\section{Introduction}
\label{intro}
In this
work  we have studied, in the framework of a nonrelativistic quark model, exclusive semileptonic and two--meson nonleptonic decays 
of the $B_c^-$
meson driven by a~ $b\to c$ or $\bar c\to \bar d,\,\bar s$ transitions
at the quark level. We have not considered semileptonic processes driven
 by the quark $b\to
u$ transition to avoid known problems both at too high $q^2$ transfers, where
one might need to include the exchange of~  $B^*_c$ resonances, and at too low
$q^2$ where recoil effects could be important~\cite{albertus}.

In order to check  the sensitivity of our results to the interquark interaction
we have used five different quark--quark potentials taken from
refs.~\cite{bhaduri,silvestre}. All the potentials used
have a confinement term plus coulomb and hyperfine terms coming from one--gluon exchange,
and differ from one another  in the power of the confining term or in the use of 
different form factors in the coulomb and hyperfine terms. Their free parameters  had been
adjusted to reproduce the light and heavy--light meson spectra.
Our central results have been obtained with the AL1 potential of 
ref.~\cite{silvestre},
while our errors show the spread of the results when using the other four potentials.

A  detailed account of the full contents of this work is now
available in ref.~\cite{hernandez}. 
\section{Semileptonic decays}
\label{sec:1}
We have made our calculations in the spectator approximation using one--body
current operators.
In table \ref{tab:1} we show our branching ratios for
semileptonic $B_c^-$ decays and compare them to the results obtained by Ivanov
{\it et
al.}~\cite{ivanov05,ivanov01}, within a relativistic quark model calculation,
and Ebert {\it et
al.}~\cite{ebert03,ebert03-2}, within the quasipotential approach to the relativistic quark
model. 
The three calculations give similar results.

\vspace*{-1cm}
\begin{table}[h!!!]
\caption{Branching ratios in \% for semileptonic $B_c^-\to c\bar c$ and 
$B_c^-\to \overline{B}$ decays. 
 In the first part of the table  $l$ stands for $l=e,\,\mu$.}
\label{tab:1}       
\begin{tabular}{llll}
\hline\noalign{\smallskip}
 & This work & \cite{ivanov05} &\cite{ebert03} \\
\noalign{\smallskip}\hline\noalign{\smallskip}
$B^-_c\to \eta_c\, l^-\,\bar{\nu}_l$ &$0.48^{+0.02}$ 
& 0.81 &0.42 \\
$B^-_c\to \eta_c\, \tau^-\,\bar{\nu}_\tau$ &$0.17^{+0.01}$ &0.22\\

$B^-_c\to \chi_{c0}\, l^-\,\bar{\nu}_l$ &$0.11^{+0.01}$&0.17\\
$B^-_c\to \chi_{c0}\,\tau^-\,\bar{\nu}_\tau$ &$0.013^{+0.001}$&
 0.013\\
$B_c^-\to J/\Psi\, l^-\,\bar{\nu}_l$ &$1.54^{+0.06}$&2.07&1.23\\
$B_c^-\to J/\Psi\,\tau^-\,\bar{\nu}_\tau$&$0.41^{+0.02}$&0.49
\\
$B_c^-\to \chi_{c1}\, l^-\,\bar{\nu}_l$ &$0.066^{+0.003}_{-0.002}$&0.092\\
$B_c^-\to \chi_{c1}\, \tau^-\,\bar{\nu}_\tau$
&$0.0072^{+0.0002}_{-0.0003}$&0.0089\\

$B_c^-\to h_c\, l^-\,\bar{\nu}_l$ &$0.17^{+0.02}$&0.27\\
$B_c^-\to h_c\, \tau^-\,\bar{\nu}_\tau$ &$0.015^{+0.001}$&0.017\\

$B_c^-\to \chi_{c2}\, l^-\,\bar{\nu}_l$&$0.13^{+0.01}$& 0.17\\
$B_c^-\to \chi_{c2}\, \tau^-\,\bar{\nu}_\tau$&$0.0093^{+0.0002}_{-0.0005}$&
 0.0082\\

$B_c^-\to \Psi (3836)\, l^-\,\bar{\nu}_l$&$0.0043_{-0.0005}$ & 0.0066\\
$B_c^-\to \Psi (3836)\, \tau^-\,\bar{\nu}_\tau$ &$0.000083_{-0.000010}$ 
&0.000099\\
\noalign{\smallskip}\hline
& This work & \cite{ivanov01}&\cite{ebert03-2}  \\
\noalign{\smallskip}\hline\noalign{\smallskip}
$B^-_c\to \overline{B}^0\, e\,\bar{\nu}_e$&$0.046^{+0.004}_{-0.007}$
&0.071&0.042\\
$B^-_c\to \overline{B}^0\, \mu\,\bar{\nu}_\mu$&$0.044^{+0.005}_{-0.006}$\\
$B^-_c\to \overline{B}_s^0\, e\,\bar{\nu}_e$&$1.06^{+0.05}_{-0.02}$
&1.10& 0.84\\
$B^-_c\to \overline{B}_s^0\, \mu\,\bar{\nu}_\mu$&$1.02^{+0.04}_{-0.02}$\\
$B^-_c\to \overline{B}^{*0}\, e\,\bar{\nu}_e$&$0.11^{+0.01}_{-0.01}$
&0.063&0.12\\
$B^-_c\to \overline{B}^{*0}\, \mu\,\bar{\nu}_\mu$&$0.11^{+0.01}_{-0.02}$\\
$B^-_c\to \overline{B}_s^{*0}\, e\,\bar{\nu}_e$&$2.35^{+0.14}_{-0.10}$
&2.37&1.75\\
$B^-_c\to \overline{B}_s^{*0}\, \mu\,\bar{\nu}_\mu$&$2.22^{+0.12}_{-0.10}$\\
\noalign{\smallskip}\hline
\end{tabular}
\end{table}

In table \ref{tab:2} we show  the
forward-backward asymmetry,  with the 
forward direction being defined by the three--momentum of the final meson, of the charged lepton measured in the leptons 
center of mass
frame.
Our results are in reasonable agreement with the ones of ref.~\cite{ivanov05} with two
exceptions corresponding to final $\chi_{c1}$ and $\Psi(3836)$ where not even the sign
of the asymmetry is the same.
\begin{table}
\caption{Forward-backward asymmetry $A_{FB}$ of the final charged lepton
($e,\,\mu$ or $\tau$)
measured in the leptons center of mass frame. }
\label{tab:2}       
\begin{tabular}{llll}
\hline\noalign{\smallskip}
&\hspace*{-.5cm}$A_{FB} (e)$
&\hspace*{-.5cm}$A_{FB} (\mu)$&\hspace*{-.5cm}$ A_{FB} (\tau) $   \\
\noalign{\smallskip}\hline\noalign{\smallskip}
$B_c\to \eta_c$\\
 This work &\hspace*{-.5cm}$0.60^{+0.01} 10^{-6}$&\hspace*{-.5cm}$0.13^{+0.01}
 10^{-1}$&\hspace*{-.5cm}$0.35$\\
\hspace*{.3cm}\cite{ivanov05}&\hspace*{-.5cm} $0.953\ 10^{-6}$&&\hspace*{-.5cm}0.36\\
$B_c\to \chi_{c0}$ \\  This work&\hspace*{-.5cm}$0.72^{+0.02} 10^{-6}$&\hspace*{-.5cm}
$0.15\ 10^{-1}$
&\hspace*{-.5cm}$0.40$\\
\hspace*{.3cm}\cite{ivanov05}& \hspace*{-.5cm}$1.31\ 10^{-6}$&&\hspace*{-.5cm}0.39\\
$B_c\to J/\Psi$ \\
 This work&\hspace*{-.5cm}$-0.19$&\hspace*{-.5cm}$-0.18_{-0.01}$ &\hspace*{-.5cm}$-0.35^{+0.02} 10^{-1}$\\

\hspace*{.3cm}\cite{ivanov05}&\hspace*{-.5cm}$-0.21$&&\hspace*{-.5cm}$-0.48\ 10^{-1}$\\
$B_c\to \chi_{c1}$ \\
This work &\hspace*{-.5cm}$-0.60_{-0.01}$&\hspace*{-.5cm}$-0.60_{-0.01}$&\hspace*{-.5cm}$-0.46$\\
\hspace*{.3cm}\cite{ivanov05}&\hspace*{-.5cm}0.19&&\hspace*{-.5cm}0.34\\
$B_c\to h_c$ \\
This work&\hspace*{-.5cm}$-0.83_{-0.05}  10^{-2}$&\hspace*{-.4cm}
$0.97_{-0.05}^{+0.01} 10^{-2}$&
\hspace*{-.3cm}$0.35$\\
\hspace*{.3cm}\cite{ivanov05}&\hspace*{-.5cm}$-3.6\ 10^{-2}$&&\hspace*{-.3cm}0.31\\

$B_c\to \chi_{c2}$ \\
 This work&\hspace*{-.5cm}$-0.14$&\hspace*{-.5cm}$-0.13$&\hspace*{-.5cm}$0.55^{+0.02} 10^{-1}$\\
\hspace*{.3cm}\cite{ivanov05}&\hspace*{-.5cm}$-0.16$&&\hspace*{-.5cm}$0.44\ 10 ^{-1}$\\
$B_c\to \Psi (3836)$\\
This work &\hspace*{-.5cm}$-0.59$&\hspace*{-.5cm}$-0.59$&\hspace*{-.5cm}$-0.42$\\
\hspace*{.3cm}\cite{ivanov05}&\hspace*{-.5cm}0.21&&\hspace*{-.5cm}0.41\\

\noalign{\smallskip}\hline
&\hspace*{-.5cm}$A_{FB} (e)$  &\hspace*{-.5cm}$A_{FB} (\mu)$  \\
\noalign{\smallskip}\hline\noalign{\smallskip}

$B_c^-\to \overline{B}^0$ \\
This work&\hspace*{-.5cm}$0.67^{+0.02} 10^{-5}$&\hspace*{-.5cm}$0.82^{+0.01} 10^{-1}$\\
$B_c^-\to \overline{B}_s^0$ \\ This work&\hspace*{-.5cm}$0.89^{+0.01} 10^{-5}$  &\hspace*{-.5cm}$0.96^{+0.01} 10^{-1}$\\
$B_c^-\to \overline{B}^{*0}$\\This work &\hspace*{-.5cm}$0.17_{-0.01}$ &\hspace*{-.5cm}$0.19_{-0.01}$ \\
$B_c^-\to \overline{B}_s^{*0}$\\ This work
&\hspace*{-.5cm}$0.14_{-0.01}$&\hspace*{-.5cm}$0.16^{+0.01}$\\
\noalign{\smallskip}\hline
\end{tabular}
\end{table}

For the decay $B_c^-\to J/\Psi\,l\bar \nu_l$ with the $J/\Psi$
 decaying into a $\mu^+\mu^-$ pair one can evaluate another angular asymmetry.
 Calling $x_\mu$ the cosine of the polar angle of the $\mu^+\mu^-$ pair,
 measured in the
 $\mu^+\mu^-$ rest frame 
 relative to the momentum of the decaying $J/\Psi$,  we have that the
 differential decay width
  $d\Gamma_{B_c^-\to\mu^+\mu^-(J/\Psi)l\bar\nu_l}/dx_\mu\equiv
 1+\alpha^*\,x_\mu$. Our results for the asymmetry parameter $\alpha^*$ are
  given in
 table~\ref{tab:3}. They are in reasonable agreement with the ones
 obtained by Ivanov {\it et al.}~\cite{ivanov05}.

 \begin{table}
\caption{$\alpha^*$ angular  asymmetry for the $\mu^+\mu^-$ pair in the
decay $B_c^-\to \mu^+\mu^- (J/\Psi)\,l\bar \nu_l$. }
\label{tab:3}       

\begin{tabular}{llll}
\hline\noalign{\smallskip}
&$\alpha^* (e)$
&$\alpha^*(\mu)$&$ \alpha^* (\tau) $   \\
\noalign{\smallskip}\hline\noalign{\smallskip}
$B_c\to J/\Psi$ \\
 This work&$-0.29_{-0.01}$&$-0.29$ &$-0.19$\\
\hspace*{.3cm}\cite{ivanov05}&$-0.34$&&$-0.24$\\
\noalign{\smallskip}\hline
\end{tabular}
\end{table}
\subsection{Heavy quark spin symmetry}
While one can not apply ordinary heavy quark symmetry to hadrons with two heavy quarks,
there is a symmetry that survives for those systems which is heavy quark spin symmetry
(HQSS).
This symmetry reflects the fact that for infinite heavy quark masses the spins of the two
heavy quarks decouple.

We have checked that our model reproduces the constraints imposed by HQSS
in the infinite heavy quark mass limit. Those constraints
relate the form factors for $B_c$ semileptonic decays into final $0^-$ and $1^-$ mesons
near the zero recoil point~\cite{jenkins93}. 

For the actual heavy quark masses  we find small deviations
from the infinite heavy quark mass limit relations for the $B_c\to\eta_c,\,J/\Psi$ case,
while corrections are large for some of the form factors in the $B_c\to\overline B^0,\overline B^{*0}$ and
$B_c\to\overline B^0_s,\overline B^{*0}_s$ cases. See ref.~\cite{hernandez} for details.

\section{Two-meson nonleptonic decays}
\begin{figure}
 \resizebox{0.5\textwidth}{!}{%
   \includegraphics{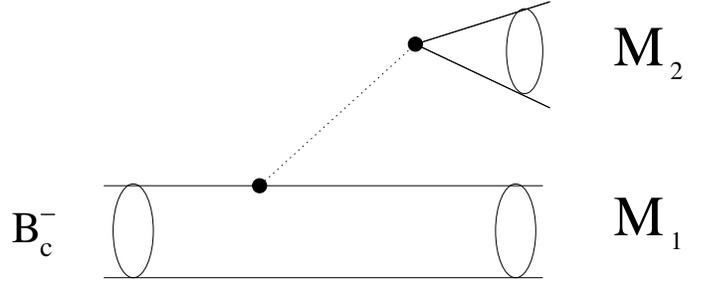}
 }
 \caption{Diagrammatic representation of $B_c^-$ two--meson nonleptonic decay in the factorization
 approximation}
 \label{fig:1}       
 \end{figure}

Neglecting  penguin contributions, the effective Hamiltonians used for 
these calculations  
are given by local four--quark operators of the current--current
type~\cite{ebert03,ivanov06}.
We work, as it is usually done, in  factorization approximation  in
which the amplitude only contains contributions as the ones depicted 
in fig.~\ref{fig:1}. To evaluate the amplitude we need different meson decay constants
that we take from experiment or lattice data. For the $\eta_c$ we use our own theoretical
result obtained with the model described in ref.~\cite{albertus05}.
%

In table~\ref{tab:4} we show results for decays that include a $c\bar c$ meson in the
final state. For decays with a~ $\pi^-,\,\rho^-,\,K^-,\,K^{*-}$ meson in the final state 
we find good agreement
with the data by Ebert {\it et al.}~\cite{ebert03} while our results are roughly a
factor of two smaller than the ones obtained by Ivanov {\it et al.}~\cite{ivanov06}.
For decays with a $D^-,\,D^{*-},\,D^-_s,\,D^{*-}_s$ meson in the final state we are in
good agreement with the results by El-Hady {\it et al.}~\cite{elhady00}, obtained using 
the Bethe--Salpeter equation,  and the ones by
Kiselev~\cite{kiselev02}, obtained within the three  point sum rules of QCD and 
nonrelativistic QCD.

In table~\ref{tab:5} we show results for decays that include a meson with a $b $ quark. Our
branching ratios for decays with a  $\overline B^0,\,\overline B^{*0}$ meson in
the final state are in very good
agreement, being $B_c^-\to \overline B^{*0}\pi^-$ the exception, with the 
results by Ebert {\it et al.}~\cite{ebert03-2}, while for the case with
a  $\overline B^0_s,\,\overline B^{*0}_s$ meson in the final state we are in very good agreement
with the results by Ivanov {\it et al.}~\cite{ivanov06}. Finally for the case
 with  a $B^-,\,B^{*-}$ meson  in the final state we find, with just one
exception ($B_c^-\to B^{*-}\pi^0$),  very good agreement with
the results by Ebert {\it et al.}~\cite{ebert03-2}.
\begin{table}
\caption{Branching ratios in \% for exclusive two--meson nonleptonic decays of the
$B_c^-$ meson that include a $c\bar c$ meson in the final state. }
\label{tab:4}       

\begin{tabular}{llll}
\hline\noalign{\smallskip}
&This work
&\cite{ivanov06} &\cite{ebert03} \\
\noalign{\smallskip}\hline\noalign{\smallskip}
$B_c^-\to\eta_{c}\,\pi^-$ &$0.094^{+0.006}$
&0.19&0.083\\
$B_c^-\to\eta_{c}\,\rho^-$ & $0.24^{+0.01}$
&0.45&0.20\\
$B_c^-\to\eta_{c}\,K^-$ &$0.0075^{+0.0005}$
&0.015&0.006\\
$B_c^-\to\eta_{c}\,K^{*-}$ &$0.013^{+0.001}$
&0.025&0.011\\
$B_c^-\to J/\Psi\,\pi^-$ &$0.076^{+0.008}$
&0.17 &0.060\\
$B_c^-\to J/\Psi\,\rho^-$ &$0.24^{+0.02}$
&0.49&0.16\\
$B_c^-\to J/\Psi\,K^-$&$0.0060^{+0.0006}$
&0.013&0.005\\
$B_c^-\to J/\Psi\,K^{*-}$ &$0.014^{+0.002}$
&0.028&0.010\\
$B_c^-\to \chi_{c0}\,\pi^-$  &$0.026^{+0.003}$
&0.055\\
$B_c^-\to \chi_{c0}\,\rho^-$ &$0.067^{+0.006}_{-0.001}$
&0.13\\
$B_c^-\to \chi_{c0}\,K^-$    &$0.0020^{+0.0002} $
&0.0042\\
$B_c^-\to \chi_{c0}\,K^{*-}$ &$0.0037^{+0.0005}$
&0.0070\\
$B_c^-\to \chi_{c1}\,\pi^-$  &$0.00014^{+0.00001}$
&0.0068\\
$B_c^-\to \chi_{c1}\,\rho^-$ &$0.010^{+0.001}_{-0.001}$
&0.029\\
$B_c^-\to \chi_{c1}\,K^-$    &$1.1^{+0.1}\,10^{-5}$
&5.1\,$10^{-4}$\\
$B_c^-\to \chi_{c1}\,K^{*-}$ &$0.00073^{+0.00007}_{-0.00002}$
&0.0018\\
$B_c^-\to h_c\,\pi^-$  &$0.053^{+0.007}$
&0.11\\
$B_c^-\to h_c\,\rho^-$ &$0.13^{+0.01}$
&0.25\\
$B_c^-\to h_c\,K^-$    &$0.0041^{+0.0006}$
&0.0083\\
$B_c^-\to h_c\,K^{*-}$ &$0.0071^{+0.0008}$
&0.013\\
$B_c^-\to \chi_{c2}\,\pi^-$  &$0.022^{+0.002}$
&0.046\\
$B_c^-\to \chi_{c2}\,\rho^-$ &$0.065^{+0.006}_{-0.002}$
&0.12\\
$B_c^-\to \chi_{c2}\,K^-$    &$0.0017^{+0.0001} $
&0.0034\\
$B_c^-\to \chi_{c2}\,K^{*-}$
&$0.0038^{+0.0003}_{-0.0002}$
&0.0065\\
$B_c^-\to \Psi(3836)\,\pi^-$  &$4.1^{+0.03}_{-0.02}\,10^{-5}$
&0.0017\\
$B_c^-\to \Psi(3836)\,\rho^-$ &$0.0020_{-0.0003}$
&0.0055\\
$B_c^-\to\Psi(3836) \,K^-$    &$3.1^{+0.2}_{-0.2}\,10^{-6}$
&0.00012\\
$B_c^-\to \Psi(3836)\,K^{*-}$ &$0.00014_{-0.00002} $
&0.00032\\

\noalign{\smallskip}\hline
\end{tabular}\\
\begin{tabular}{lllll}
&This work
&\cite{ivanov06} &\cite{elhady00}&\cite{kiselev02} \\
\noalign{\smallskip}\hline\noalign{\smallskip}
$B_c^-\to\eta_{c}\,D^-$ & $0.014^{+0.001}$
&0.019&0.014&0.015\\
$B_c^-\to\eta_{c}\,D^{*-}$ &
$0.012^{+0.001}$
&0.019&0.013&0.010\\
$B_c^-\to J/\Psi\,D^-$ &
$0.0083^{+0.0005}$
&0.015&0.009&0.009\\
$B_c^-\to J/\Psi\,D^{*-}$ & $0.031^{+0.001}$
&0.045&0.028&0.028\\
$B_c^-\to\eta_{c}\,D_s^-$ & $0.44^{+0.02}$
&0.44&0.26&0.28\\
$B_c^-\to\eta_{c}\,D_s^{*-}$ & $0.24^{+0.02}$
&0.37&0.24&0.27\\
$B_c^-\to J/\Psi\,D_s^-$ & $0.24^{+0.02}$
&0.34&0.15&0.17\\
$B_c^-\to J/\Psi\,D_s^{*-}$ & $0.68^{+0.03}$
&0.97&0.55&0.67\\

\noalign{\smallskip}\hline
\end{tabular}
\end{table}

\begin{table}
\caption{Branching ratios in \% for exclusive two--meson nonleptonic decays of the
$B_c^-$ meson that include a $\overline B,\,B$ meson in the final state. }
\label{tab:5}       

\begin{tabular}{llll}
\hline\noalign{\smallskip}
&This work
&\cite{ivanov06} &\cite{ebert03-2} \\
\noalign{\smallskip}\hline\noalign{\smallskip}
$B_c^-\to\overline{B}^0\,\pi^-$ & $0.11^{+0.01}_{-0.01}$
&0.20&0.10\\
$B_c^-\to\overline{B}^0\,\rho^-$ & $0.14^{+0.02}_{-0.02}$
&0.20&0.13\\
$B_c^-\to\overline{B}^0\,K^-$ & $0.010^{+0.001}_{-0.001}$
&0.015&0.009\\
$B_c^-\to\overline{B}^0\,K^{*-}$ &
$0.0039^{+0.0003}_{-0.0005}$
&0.0048&0.004\\
$B_c^-\to \overline{B}^{*0}\,\pi^-$ &$0.072^{+0.012}_{-0.012}$
&0.057&0.026\\
$B_c^-\to \overline{B}^{*0}\,\rho^-$ &$0.58^{+0.05}_{-0.08}$
&0.30&0.67\\
$B_c^-\to \overline{B}^{*0}\,K^-$
&$0.0048^{+0.0007}_{-0.0008}$
&0.0036&0.004\\
$B_c^-\to \overline{B}^{*0}\,K^{*-}$
&$0.030^{+0.002}_{-0.004}$
&0.013&0.032\\
$B_c^-\to\overline{B}_s^0\,\pi^-$ & $3.51^{+0.19}_{-0.06}$
&3.9&2.46\\
$B_c^-\to\overline{B}_s^0\,\rho^-$ & $2.34^{+0.05}_{-0.06}$
&2.3&1.38\\
$B_c^-\to\overline{B}_s^0\,K^-$ & $0.29^{+0.01}_{-0.01}$
&0.29&0.21\\
$B_c^-\to\overline{B}_s^0\,K^{*-}$ & $0.013_{-0.001}$
&0.011&0.0030\\
$B_c^-\to \overline{B}_s^{*0}\,\pi^-$ &$2.34^{+0.19}_{-0.14}$
&2.1&1.58\\
$B_c^-\to \overline{B}_s^{*0}\,\rho^-$ &$13.4^{+0.5}_{-0.6}$
&11&10.8\\
$B_c^-\to \overline{B}_s^{*0}\,K^-$ &$0.13^{+0.01}_{-0.01}$
&0.13&0.11\\

\noalign{\smallskip}\hline
\end{tabular}\\
\begin{tabular}{llll}
 & This work&\cite{ivanov06}&\cite{ebert03-2} \\
\noalign{\smallskip}\hline\noalign{\smallskip}
$B_c^-\to{B}^-\,\pi^0$ & $0.0038^{+0.0005}_{-0.0006}$
&0.007&0.004\\
$B_c^-\to{B}^-\,\rho^0$ & $0.0050^{+0.0004}_{-0.0007}$
&0.0071&0.005\\
$B_c^-\to{B}^-\,K^0$ & $0.25^{+0.03}_{-0.04}$
&0.38  &0.24\\
$B_c^-\to{B}^-\,K^{*0}$ &$0.093^{+0.006}_{-0.013}$
&0.11 &0.09 \\
$B_c\to {B}^{*-}\,\pi^0$ &$0.0025^{+0.0004}_{-0.0005}$
&0.0020&0.001\\
$B_c\to {B}^{*-}\,\rho^0$ &$0.020^{+0.002}_{-0.003}$
&0.011 &0.024\\
$B_c\to {B}^{*-}\,K^0$&$0.12^{+0.02}_{-0.02}$
&0.088 &0.11\\
$B_c\to {B}^{*-}\,K^{*0}$&$0.73^{+0.06}_{-0.10}$
&0.32 &0.84 \\

\noalign{\smallskip}\hline
\end{tabular}\end{table}
%

%
%

\begin{acknowledgement}
 This research was supported by DGI and FEDER funds, under contracts
FIS2005-00810, BFM2003-00856 and FPA2004-05616, by Junta de
Andaluc\'\i a and Junta de Castilla y Le\'on under contracts FQM0225
and SA104/04, and it is part of the EU integrated infrastructure
initiative Hadron Physics Project under contract number
RII3-CT-2004-506078.  J. M. V.-V. acknowledges a contract E.P.I.F. with the
University of Salamanca.
\end{acknowledgement}

\end{document}